\documentclass[a4paper,11pt]{article}
\usepackage[dvipsnames,table]{xcolor}
\usepackage{amsfonts,amsthm,amssymb}
\usepackage{dsfont}
\usepackage{booktabs}
\usepackage{tikz}
\usepackage{setspace}
\usepackage{subfigure}
\usepackage{floatrow}
\usepackage[flushleft]{threeparttable}
\usepackage{endnotes}
\usepackage{multirow}
\usepackage{lipsum}
\usepackage{graphicx}
\usepackage{epstopdf}
\usepackage{booktabs}
\usepackage{siunitx}
\usepackage{threeparttable}
\usepackage{geometry}
\usepackage{lineno}
\usepackage{bm}
\usepackage{mathtools}
\mathtoolsset{showonlyrefs=true}

\usepackage[authoryear]{natbib}

\usepackage[plainpages=false, pdfpagelabels]{hyperref} 
\hypersetup{
	colorlinks   = true,
	citecolor    = RoyalBlue, 
	linkcolor    = blue, 
	urlcolor     = blue
}

\newcommand\COMP{\hbox{C\kern -.58em {\raise .54ex \hbox{$\scriptscriptstyle |$}}
\kern-.55em {\raise .53ex \hbox{$\scriptscriptstyle |$}} }}
\newcommand\NN{\hbox{I\kern-.2em\hbox{N}}}
\newcommand\RR{\hbox{I\kern-.2em\hbox{R}}}
\newcommand\sRR{{\it \hbox{I\kern-.2em\hbox{R}}}}
\newcommand\QQ{\hbox{I\kern-.53em\hbox{Q}}}
\newcommand\PP{\hbox{I\kern-.53em\hbox{P}}}
\newcommand\EE{\hbox{I\kern-.53em\hbox{E}}}
\newcommand\ZZ{{{\rm Z}\kern-.28em{\rm Z}}}
\newcommand\be{\begin{equation}}
\newcommand\ee{\end{equation}}

\newcommand\beq{\begin{eqnarray}}
\newcommand\eeq{\end{eqnarray}}
\newcommand\bq{\begin{eqnarray*}}
\newcommand\eq{\end{eqnarray*}}

\numberwithin{equation}{section}

\allowdisplaybreaks

\begin{document}
\title{Liquidation, Leverage and Optimal Margin in Bitcoin Futures Markets}
\author{
	Zhiyong Cheng$^{a}$,\ \
	Jun Deng$^{a}$
	\footnote{Corresponding Author. School of Banking and Finance, University of International Business and Economics, Beijing, China, 100029. Email: jundeng@uibe.edu.cn},\ \
	Tianyi Wang$^{a}$,\ \
	Mei Yu$^{a}$
}

\date{}

\maketitle
%

\begin{abstract}
 Using the generalized extreme value theory to characterize tail distributions, we address liquidation, leverage, and optimal margins for bitcoin long and short futures positions. The empirical analysis of perpetual bitcoin futures on  BitMEX shows that (1)  daily forced liquidations to outstanding futures are substantial at  3.51\%, and  1.89\%   for long and short; (2) investors   got forced liquidation do trade aggressively with average leverage of  60X; and (3) exchanges should elevate current   1\% margin requirement  to  33\% (3X leverage) for long and  20\% (5X leverage) for short to reduce the daily margin call probability to 1\%. Our results further suggest normality assumption on return significantly underestimates optimal margins. Policy implications are also discussed. 
\end{abstract}


\vskip 3cm 

%
\clearpage \newpage

\section{Introduction}
 {The largest cryptocurrency, bitcoin,  accounts for more than 70\% of the total market capitalization reported by  \href{http://coinmarketcap.com }{CoinMarketCap} on 14 December 2020}. Compared with other traditional assets, bitcoin price is  {more volatile}: 30-day volatility reaches to 167.24\% on 31 March, 2020\footnote{See \href{https://www.forbes.com/sites/cbovaird/2020/04/08/bitcoin-volatility-reached-a-6-year-high-in-march/?sh=4dca9cbac211}{Forbes report}. {The highest 30-day volatility of the S\&P 500 index so far is 89.53\% on 24 October 2008.}}. This extraordinarily high price volatility imposes a tremendous risk to various market participants, see \cite{chaim2018volatility}, \cite{alexander2021Optimal},  \cite{jun2020minimum} and \cite{scaillet2020high}. 

Futures contracts are commonly used to hedge spot price risk. We refer this   rich field   to \cite{figlewski1984margins}, \cite{daskalaki2016effects} and \cite{alexander2019parsimonious} for traditional equity,  commodity and currency markets. For  bitcoin futures markets, the hedge effectiveness and improvement of portfolio performance are well studied in \cite{alexander2020bitmex}, \cite{sebastiao2020bitcoin}, \cite{deng2019optimal} and others.  By November 2020, bitcoin futures  monthly trading
volume   reached to \$1.32 trillion notional, reported by \href{https://www.cryptocompare.com/media/37621821/cryptocompare_exchange_review_2020_11.pdf}{CryptoCompare}.   

\begin{figure}[htpb]
	\centering
	\begin{minipage}[b]{0.85\textwidth}
		\includegraphics[width=1\textwidth]{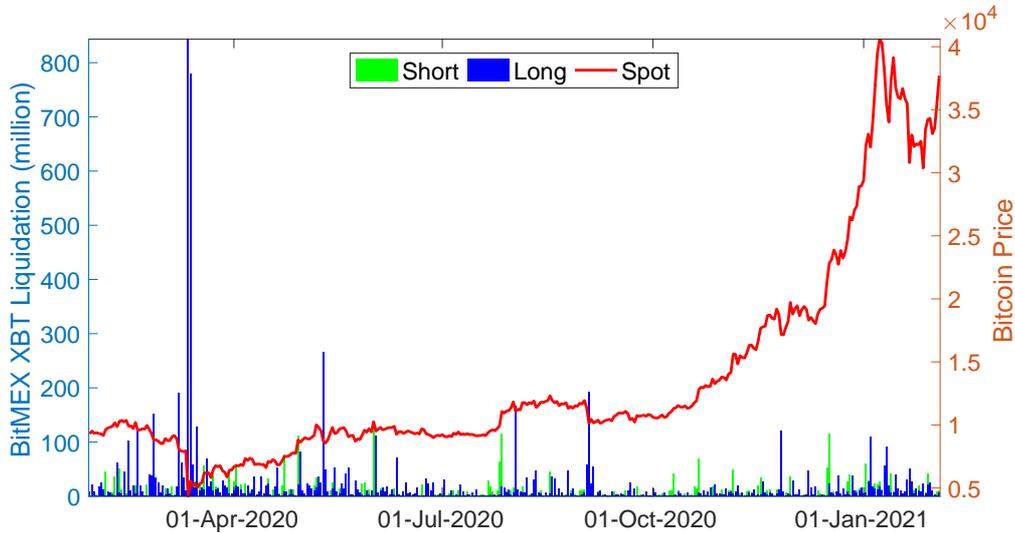}
	\end{minipage}
	\caption{Daily Liquidation  Volumes of BitMEX Perpetual Futures}
	\label{Liquidation}
	\floatfoot{Note: The green (blue) bar plots the forced liquidation of long (short) bitcoin perpetual futures on BitMEX from Jan. 2020 to Feb. 2021. On ``Black Thursday'' March 12, 2020, the long position's daily forced liquidation reached 843 million USD. The maximal short liquidation 132 million  occurred on June 1, 2020 when the price rise over 10\%.  The average daily liquidations are around 20 million USD and 10 million USD  for long and short. The right y-axis corresponds to the daily bitcoin spot price (red line). The data is manually collected from \href{https://coinalyze.net/coinalyze.net}{coinalyze}.}
\end{figure}

In this paper, our leitmotif to study optimal margins stems from several facts observed in bitcoin futures markets. First, as is well-known, the margin requirement is one of the key market designs by exchanges to maintain the integrity, liquidity, and efficiency of futures markets, see  \cite{figlewski1984margins}.  However, to the best of our knowledge, existing researches on bitcoin futures' margin are limited. One exception is the recent work of \cite{alexander2021Optimal} that investigates the optimal bitcoin futures hedge under investors' margin constraints and default aversion.  Second,  bitcoin futures contracts are traded across several exchanges,    from regulated exchanges such as CME and Bakkt to less-regulated online exchanges such as BitMEX, Binance, Huobi, and OKEx.  These online exchanges constitute a competitive marketplace that attracts a wider range of participants and dominates bitcoin futures trading volume.  Third,   leverages allowed by those exchanges differ tremendously,   from 2X of regulated to 100X of less-regulated. The unconventional high leverage of those less-regulated exchanges induced substantial forced liquidations that shaped the bitcoin futures market's evolution. For instance,  Binance overtakes BitMEX as the most liquid perpetual futures market after the ``Black Thursday'' 12 March 2020.   On that day, the bitcoin price fell by 50\%, and the daily forced liquidation of the long position reached to highest 843 million USD\footnote{See  \href{https://www.tronweekly.com/bitmexs-bitcoin-market-share-lose-to-binance-significant-after-black-thursday/}{Tronweekly} and \href{https://coingape.com/binance-overtakes-bitmex-as-bitcoins-most-liquid-perpetual-swap/}{Coingape} reports.}.      Figure \ref{Liquidation} plots daily forced liquidation on BitMEX from Jan. 2020 to Feb. 2021, where the average daily liquidation is approximate   20 million USD for long and  10 million USD for short.

This article contributes to the literature in several ways.  First, by using daily forced liquidation data of BitMEX, we highlight the speculative activity,  aggressiveness and risk preference of market participants by estimating the speculation index,  the average percentage of liquidation and average leverages used in long and short positions.  Second, we show that generalized extreme value theory (GEV) can effectively capture the bitcoin futures price's fat-tail feature\footnote{GEV distribution is  also used  for margin settings in   \cite{longin1999optimal}, \cite{dewachter1999setting}, \cite{cotter2001margin}, and more recent work of \cite{gkillas2018application}.}.  Finally, we give optimal margins for long and short positions using GEV and market data that provide guidance for exchanges to design bitcoin futures specifications, especially margin requirements.  

With bitcoin perpetual futures data from BitMEX,  our empirical analysis uncovered several interesting aspects of bitcoin futures markets. First, The average  speculation index sits at 3.75 which is much larger than S\&P 500 (0.15), Nikkei (0.21) and DAX (0.45). This shows the extreme speculation activity in BitMEX perpetual market.   Second, the daily average percentage of forced liquidation for long and short positions are substantial at levels of  3.51\% and  1.89\%, respectively. Despite those statistics, the average leverage of those  got forced liquidation is similar (around 60X) for both positions. That suggests some participants do trade aggressively and utilize high leverages regardless of high price risk. Third, assuming 1\% daily margin call probability, the optimal margins are approximate  33\% (3X leverage) for long position and 20\% (5X leverage) for short. At last, the GEV distribution is critical for accurately estimating those margins as the normal distribution assumption on return significantly underestimates margin levels by at least 50\%.

We organize the rest of the paper as follows.  Section \ref{2}  introduces bitcoin perpetual futures and generalized extreme value theory. Empirical analysis using BitMEX perpetual futures is conducted in Section \ref{3}. Section \ref{sec_speculation} studies speculation, leverage and liquidation on BitMEX.  Section \ref{4} concludes the paper and an appendix collects tables and figures.

\section{Bitcoin Perpetual Futures and Extreme Value Theory}\label{2}

\subsection{Bitcoin Perpetual Futures}\label{2.1}
Two types of futures contracts are traded across exchanges: standard futures on CME and Bakkt and perpetual futures (perpetuals) without expiry date on BitMEX, Binance, and Huobi, to name a few. Perpetual futures depart from standard futures in two aspects. Firstly, bitcoin perpetuals are quoted in US dollars (USD) and settled in Bitcoin  (XBT); while bitcoin standard futures are both quoted and settled in USD\footnote{It resembles quanto-options that are settled in domestic currency but quoted in foreign currency. }. For instance, each CME bitcoin standard futures contract has a notional value of 5 XBT, while one BitMEX bitcoin perpetual futures has a notional value of 1 USD. Secondly, the bitcoin margin deposit is required for trading perpetuals; while the USD margin is used for standard futures. Since perpetuals account for around 90\% trading volume, we only focus on perpetuals here\footnote{For more discussion on perpetuals in hedge, we refer to \cite{alexander2021Optimal}.}. 
 	
Specifically, suppose an investor who enters into  one long position of bitcoin perpetual futures with a fixed notional amount of $\Pi$ USD at time $t_1$, and   closes her position at later time $t_2$.  The perpetuals prices, denominated in USD,  changed from  $F_{t_1}$  at time $t_1$ to   $F_{t_2}$ at time $t_2$. This is equivalent to the change from $\frac{\Pi}{F_{t_1}}$ XBT to $\frac{\Pi}{F_{t_2}}$ XBT  for each perpetuals.   Then, the realized pay-off in XBT is given by the difference between the ``enter value'' and ``exit value'':
\begin{align}\label{payoff-long}
	\textit{Long Pay-off} = \frac{\Pi}{F_{t_1}} - \frac{\Pi}{F_{t_2}}.
\end{align}
 As a result, the pay-off is \textit{inversely} related to quotes and perpetuals are also called ``inverse'' futures.  {The gain increases when futures price increases as standard futures. However, the magnitude is significantly different from the latter\footnote{ {For standard futures, the long gain equals $F_{t_2}-F_{t_1}$}.}. Similarly, for the short position, the pay-off in XBT is}
\begin{align}\label{payoff-short}
	\textit{Short Pay-off} =  \frac{\Pi}{F_{t_2}} - \frac{\Pi}{F_{t_1}}.
\end{align}
Implied by equations \eqref{payoff-long} and \eqref{payoff-short},  long positions incur  greater loss than    short positions for the same magnitude of price change $\Delta F$. As a result, margins for long and short would be  asymmetric even the perpetuals price ($F_t$) fluctuates symmetrically. Empirical results in Section \ref{3} confirm the asymmetry.

\subsection{Extreme Value Theory and Optimal Margin}\label{2.2}
Let $\Delta F_{t_1},\Delta F_{t_2},\cdots,\Delta F_{t_n}$ be {quote} price changes observed on discrete time $t_i=1,2,\cdots, n$.  The  maximum and minimum changes over $n$ periods are $\textbf{Max}(n)= \max(\Delta F_{t_1}, \Delta F_{t_2}, \cdots, \Delta F_{t_n})$ and $\textbf{Min}(n)=  \min(\Delta F_{t_1}, \Delta F_{t_2}, \cdots, \Delta F_{t_n})$, respectively. For a wide range of possible distributions of price changes $(\Delta F_{t_i})$, the limit distributions of $\textbf{Max}(n)$ and $\textbf{Min}(n)$ follow  the generalized extreme value distribution (GEV), see \cite{jenkinson1955frequency}. The extreme distribution is characterized by location parameter $\mu$, scale parameter $\sigma$, and tail parameter  $\tau$.  Due to possible asymmetry in left and right tails, GEV parameters could be different for $\textbf{Max}(n)$ and $\textbf{Min}(n)$ in equations \eqref{max_limiting} and \eqref{min_limiting}.    
\begin{align}\label{max_limiting}
	G_{max}(x) &=\mathrm{exp}\left[-\left(1- \frac{x-\mu^{max}}{\sigma^{max}} \cdot \tau^{max}\right)^{1/\tau^{max}}\right],
	\quad \mbox{and} \\
	G_{min}(x)&=1-\mathrm{exp}\left[-\left(1+\frac{x-\mu^{min}}{\sigma^{min}} \cdot \tau^{min}\right)^{1/\tau^{min}}\right].  \label{min_limiting}
\end{align}
The introduction of GEV to study optimal margin can be traced back to \cite{longin1999optimal}, which focuses on silver futures traded on COMEX. Recently, \cite{gkillas2018application} also used GEV to estimate Value-at-Risk and expected shortfall of several cryptocurrencies. 

\begin{figure}[htbp]
	\centering
	\begin{minipage}[b]{1\textwidth}
		\includegraphics[width=1\textwidth]{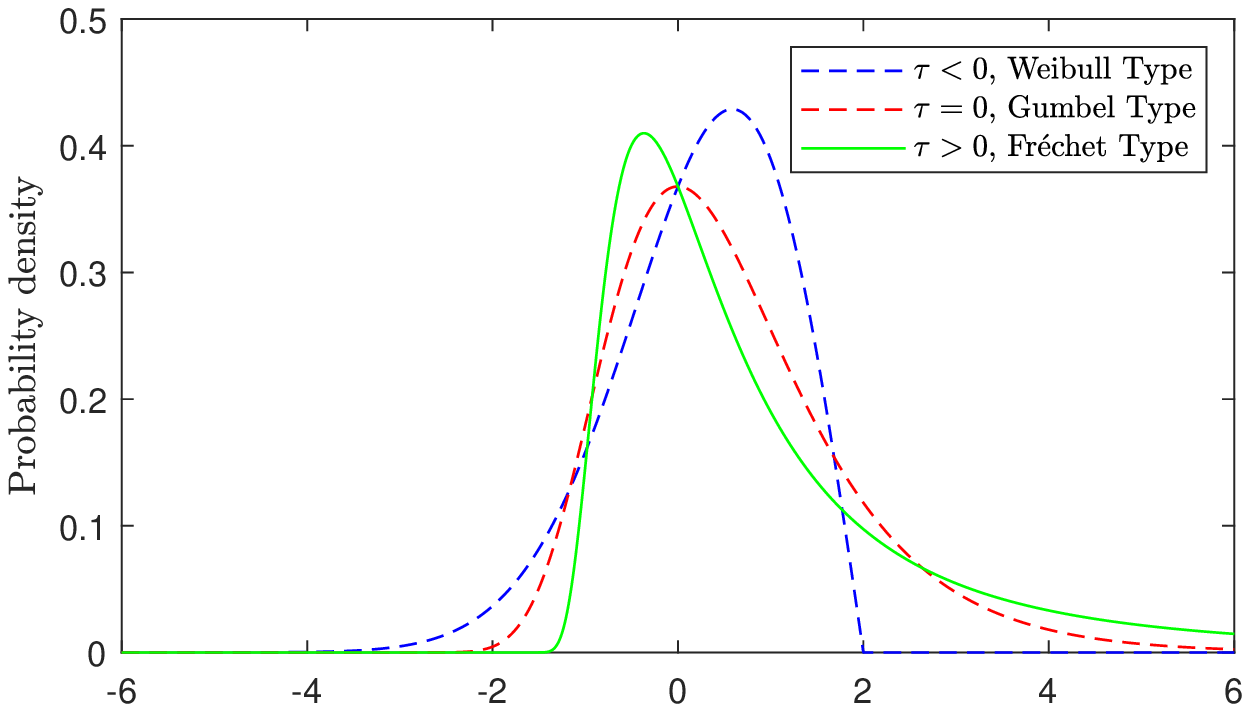}
	\end{minipage}
	\caption{Gumbel, Weibull, and Fr$\acute{e}$chet Distributions}
	\label{function}
	\floatfoot{Note: We consider three tail parameters $\tau=0.5, 0, -0.5$. The Fr$\acute{e}$chet distribution (green line) is fat-tailed as its tail is slowly decreasing; the Gumbel distribution (red line) is thin-tailed as its tail is rapidly decreasing; and the Weibull (blue line) has no tail after a certain point.}
\end{figure}

Depending on the  tail parameter $\tau$, extreme distributions fall into three categories: Gumbel ($\tau=0$), Weibull ($\tau <0$), and Fr$\acute{e}$chet ($\tau>0$) distributions, which are illustrated in Figure \ref{function}. 
\begin{center}
	$<$Insert Figure \ref{function} here.$>$
\end{center}

Over  $n$ periods, for a long position, the margin call is triggered once the minimum price change $\textbf{Min}(n)$  {is lower than} margin deposit $\textbf{MD}_{long}$.  Once    triggered, the long position would be automatically  liquidated  if the investor could not pull up enough bitcoin deposit.   {The corresponding margin call probability,} $\textbf{p}^{long}$, is given by 
\begin{align}\label{long}
	\textbf{p}^{long}=\mathbb{P}\mbox{rob}\left(\textbf{Min}(n)<-\textbf{MD}_{long}\right)\simeq G_{min}(-\textbf{MD}_{long}).
\end{align}
In parallel, for the short position, we have 
\begin{align}\label{short}
	\textbf{p}^{short}=\mathbb{P}\mbox{rob}\left(\textbf{Max}(n)>\textbf{MD}_{short}\right)\simeq 1-G_{max}(\textbf{MD}_{short}).
\end{align} 
Using   limiting distributions in  \eqref{max_limiting} and \eqref{min_limiting}, given acceptable margin call probabilities    $\textbf{p}^{long}$ and $\textbf{p}^{short}$,     minimal margins    $\textbf{MD}_{long}$ and $\textbf{MD}_{short}$  can be analytically expressed   as
\begin{align}\label{long-margin-gev}
	\textbf{MD}_{long} &= -\mu^{min} +\frac{\sigma^{min}}{\tau^{min}}\left[1-\left(-\ln\left(1-\textbf{p}^{long}\right)\right)^{\tau^{min}}\right],  \qquad\mbox{and}\\
	\textbf{MD}_{short} &= \mu^{max} +\frac{\sigma^{max}}{\tau^{max}}\left[1-\left(-\ln\left(1-\textbf{p}^{short}\right)\right)^{\tau^{max}}\right]. \label{short-margin-gev}
\end{align}
In Section \ref{3} below, we conduct empirical analysis to estimate left and right GEV distribution parameters $(\tau^{min},\mu^{min}, \sigma^{min})$ and $(\tau^{max},\mu^{max}, \sigma^{max})$ and related long and short margin levels $\textbf{MD}_{long}$ and $\textbf{MD}_{short}$.

\section{Data and Empirical Analysis}\label{3}
\subsection{Data Description}\label{3.1}
The BitMEX exchange, founded in 2014, is the largest online platform for trading bitcoin futures, which  runs continuously 24/7 a week and offers various cryptocurrency futures contracts, such as Bitcoin and Ethereum. On BitMEX, both perpetuals and fixed expiration futures are listed.  {The perpetual futures contracts dominate the market by contributing 90\% of the total trading volume.}  Perpetuals traded on BitMEX have notional value $\Pi=1$ USD and the maximum leverage allowed is 100X  {(margin requirement is 1\%)}.

To investigate the average leverage used by market participants, we manually collect the daily volume,  forced liquidation, open interest and daily OHLC prices (open, high, low, close) data from \href{coinalyze.net}{Coinalyze}. The data spans from January 29, 2020 to February 3, 2021 and contains 372 entries\footnote{As BitMEX runs 24/7, the open (close) price for each day  is defined as the first (last) trade   at 0:00 UTC and 24:00 UTC. Since liquidation data is seldomly released, we manually collect the earliest of BitMEX from \href{coinalyze.net}{Coinalyze}.}. Besides, we also download 5min BitMEX perpetual futures   price data using the API provided by  \href{https://www.bitmex.com/}{BitMEX} to investigate   optimal margins. The BitMEX perpetual data-set consists of 431, 346 observations from January 1, 2017 to February 6, 2021.   

Following \cite{longin1999optimal},   price changes for perpetual futures are defined in percentage changes $\Delta \widehat{F}_{t}=  \frac{\frac{1}{F_{t-1}} - \frac{1}{F_{t}}}{\frac{1}{F_{t-1}}} =  1 - \frac{{F}_{t-1}}{{F}_{t}}$. 
This definition has the  advantage of being a stationary time series and  num\'eraire  independent. Margin level is in  
  terms of percentage. Its left (right) tail is associated with the loss of a long (short) position. We also define an alternative price change as $\Delta  {F}_{t} = \frac{F_t - F_{t-1}}{F_{t-1}} =  \frac{{F}_{t}}{{F}_{t-1}} - 1$ if the investor trades USD settled standard bitcoin futures. The difference in optimal margins under two definitions highlights the impact of perpetuals' inverse pay-off structure.

\begin{center}
	$<$Insert Table \ref{bitmex_summary} here.$>$
\end{center}

\begin{figure}[htbp]
	\centering
	\begin{minipage}[b]{1\textwidth}
		\includegraphics[width=1\textwidth]{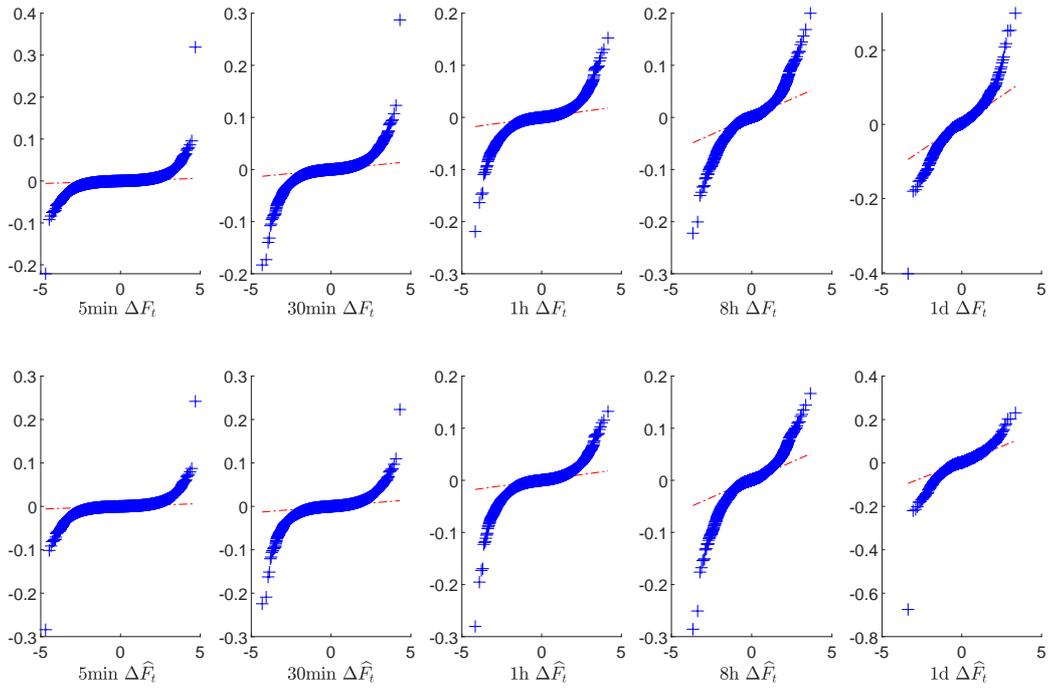}
	\end{minipage}
	\caption{Q-Q Plot of   Price Changes $\Delta F$ and $\Delta \widehat{F}$ on BitMEX}
	\label{QQ}
\end{figure}

Table \ref{bitmex_summary} reports the summary statistics of price changes  $\Delta F$ and $\Delta \widehat{F}$, sampled at different frequencies.  On daily frequency,   both price changes are negatively skewed and leptokurtic, and the magnitude for perpetuals   is higher, evidenced by a more negative skewness (-3.67 vs. -0.24). The price change distribution for perpetuals favors the short position than the long position as $\Delta \widehat{F}$ has a lower minimum (-82.67\% vs. -45.26\%), a lower maximum (22.00\% vs. 28.20\%) and, as a result, a lower mean (0.14\% vs. 0.35\%) as well as a thicker tail\footnote{ {The kurtosis  65.36 for $\Delta \widehat{F}$ and 14.10 for $\Delta F$}.}. These facts are also confirmed by the Q-Q plot in Figure \ref{QQ}.

\begin{center}
	$<$Insert Figure \ref{QQ} here.$>$
\end{center}

\subsection{GEV Parameter Estimation}\label{3.3}
As mentioned above, the left/right tail is directly linked to the loss of long/short position. Two sets of parameters $(\mu,\sigma,\tau)$ are estimated separately.

As there is only one extreme value for the full sample, estimating GEV distribution parameters often relies on the block extreme technique. Here, we follow the estimation procedure of \cite{longin1999optimal}.  The technique divides the whole sample into non-overlapping sub-samples containing $n$ observations, and each sub-sample provides us a maximum change and a minimum change. For  the $i$-th block $\{\Delta X_{(i-1)n+1},\Delta X_{(i-1)n+2},\cdots,\Delta X_{in}\}$ ($X = \widehat F$ and $F$, respectively), we denote the block maximum and minimum as $\min_i$ and $\max_i$. If we have a total of $N$ observations, the block extreme sets $\{\min_1,...,\min_{[N/n]}\}$ and $\{\max_1,...,\max_{[N/n]}\}$ are used to estimate GEV parameters\footnote{$[N/n]$ is the integer part of $N/n$}. For different sampling frequencies, the block size $n$ varies. In particular, the block spans 8/24/48 hours for 5/30/60 min sampling and 5/10 days for 8/24 hours sampling frequencies\footnote{ {Our empirical results are robust to the block size. The 8 hours are selected as BitMEX charges funding rate for investors holding positions every 8 hours.}}, and we point out the sampling frequency could be treated as the screening-frequency that exchanges/investors monitor the futures positions. We report the corresponding results in Table \ref{para-bitmex}.

The tail parameter $\tau$ is consistently positive, indicating extreme distributions are all Fr$\acute{e}$chet types. The magnitude of $\tau$ increases as the sampling frequency increases, which means the tail is thicker for high-frequency returns. On the other hand, the location and scale parameters are higher for low-frequency returns as extreme price changes increase for longer horizons. Slight differences are observed for left and right tail parameters of standard futures $\Delta F$. However, for perpetual futures $\Delta \widehat{F}_{t}$, the tail parameter $\tau$'s differ significantly for left and right tails, especially for low-frequency cases such as 8h (8 hours) and 1d (1 day). For daily return, the left tail parameter $\tau$ is more than double of the right, and this indicates a fatter left tail (higher risk for the long position).

\begin{center}
	$<$Insert Table \ref{para-bitmex} here.$>$
\end{center}

\begin{figure}[htbp]
	\centering
	\begin{minipage}[b]{1.1\textwidth} 
		  \includegraphics[width=1\textwidth]{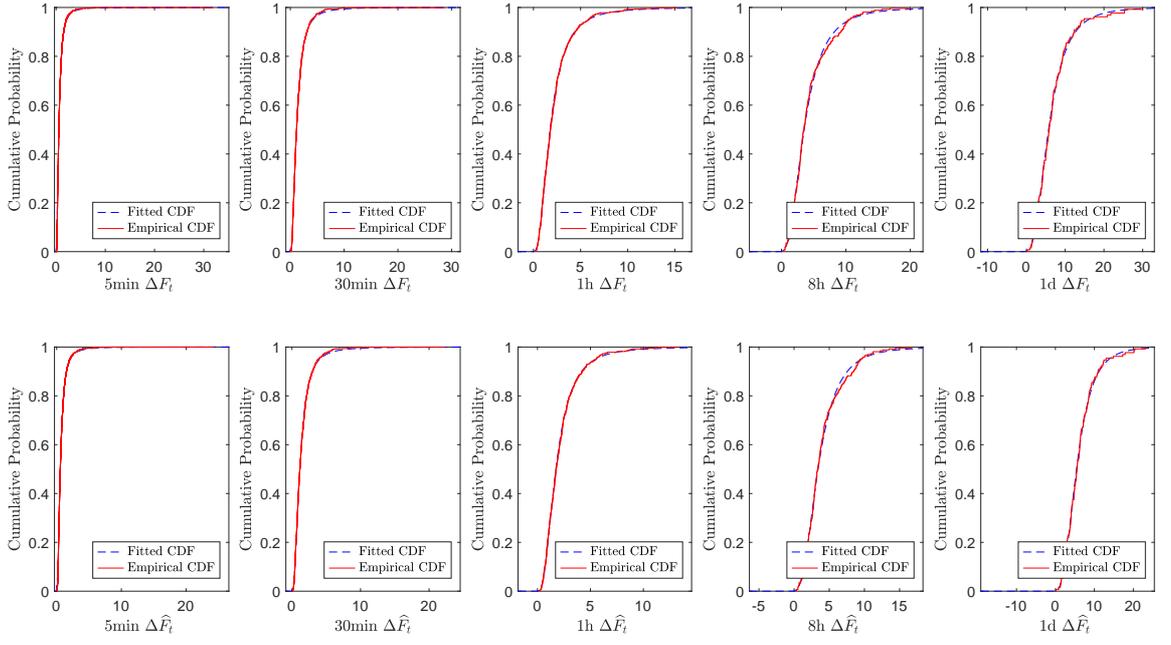}
	\end{minipage}
	\caption{Empirical and Fitted Right Tail CDF of $\Delta F_{t}$ and $\Delta \widehat{F}_{t}$}
	\label{gevfit}
	\floatfoot{Note: This figure shows the empirical and fitted CDF of $\Delta F_{t}$ and $\Delta \widehat{F}_{t}$ under 5min, 30min, 1h, 8h, and 1d observation frequencies. All the fitted right tail CDF's are computed by \eqref{max_limiting} with parameters in Panel A of Table \ref{para-bitmex}.}
\end{figure}

Finally, we plot the empirical and fitted cumulative distribution functions (CDFs) for both $\Delta F_{t}$ and $\Delta \widehat{F}_{t}$ in Figure \ref{gevfit}. Results suggest that the GEV method is suitable for capturing tail behaviors and investigating optimal margins for bitcoin futures.

\begin{center}
	$<$Insert Figure \ref{gevfit} here.$>$
\end{center}

\subsection{Optimal Margin}\label{3.4}
In this section, using estimated parameters in Table \ref{para-bitmex} of Section \ref{3.3}, we calculate optimal margins via \eqref{long-margin-gev} and \eqref{short-margin-gev} and present empirical results in Table \ref{oml-bitmex}.
Four margin call probabilities and five holding periods are discussed. Both perpetual futures and standard futures are studied to illustrate the impact of the inverse feature on margin requirements. The results for standard futures also provide guidelines for bitcoin futures traded on CME and Bakkt. Moreover, to match the current margin requirements set by exchanges, we also calculate the one common optimal margin level for both positions by simultaneously encompassing the left and right tails. For each case, we also provide the optimal margin based on the normal distribution assumption (in parentheses) to show the importance of GEV distribution.

\begin{center}
	$<$Insert Table \ref{oml-bitmex} here.$>$
\end{center}

For each holding period, as expected, the optimal margin increases with the decreasing margin call tolerance. For standard futures, the margin increases from 7.86\% (7.66\%) to 32.95\% (35.93\%) for short (long) position when margin call probability varies from $p=0.1$ to $p=0.001$; while perpetual futures margin increases from 7.26\% (8.36\%) to 26.34\% (46.98\%) accordingly for 8 hours holding period. 

The difference in optimal margin levels for different positions is smaller when the margin call tolerance is higher. For different kinds of futures, the standard one has a rather balanced margin requirement for both positions, while the perpetuals require a higher margin for long than short, especially for longer holding periods and lower tolerance of margin calls. The suggested margin for 8 hours holding period is almost doubled for the long than the short position (46.98\% vs. 26.34\%) when the margin call probability is limited to 0.001. Such \textit{asymmetric effect} is the direct result of the difference in tail parameters listed in Table \ref{para-bitmex}.

To match the exchanges' practice of setting equal margins for both positions, we pool maximums and minimums together as $\{-\min_1,...,-\min_{[N/n]},\max_1,...,\max_{[N/n]}\}$ and estimate one set of parameters and the corresponding margin levels in Panel C of Table \ref{oml-bitmex}. As expected, the common margin level lies somewhere between the long and short cases. For example, for 8 hours holding period with call probability $p=0.001$, the common margin level for perpetual futures is 34.81\%, whereas it is  46.98\% for the long position and  26.34\% for the short position. Finally, the corresponding optimal margins obtained under the normality assumption of return would significantly underestimate optimal margins by at least half.

In the following section, we quantify the speculation activity, trading aggressiveness and liquidation on BitMEX. It reflects market participants' risk preference in this highly volatile market. 

\section{Speculation, Leverage and Liquidation on BitMEX}\label{sec_speculation}
Previously, we estimated optimal margins for both long and short positions under different monitoring frequencies by using perpetual price data. However,  to get a sense of investors' trading activity and risk preference in the bitcoin futures market, calculating speculation percentage and precise leverages used by investors requires traders' account-level data, which is unavailable from the exchange. Fortunately, \href{coinalyze.net}{Coinalyze} reports the forced liquidation of both long and short. Combined with the daily OHLC prices (open, high, low, close) of bitcoin perpetual futures and trading volume and open interest, we can roughly estimate speculation activity and  the leverage used by investors who experienced forced liquidation within a particular day.

First, following \cite{garcia1986lead},  \cite{lucia2010measuring}, and \cite{kim2015does},  we define the speculation index as the ratio  of   daily trading volume to daily open interest, i.e.
\begin{align}\label{speculation_index}
	\cal{SI} = \frac{\mbox{trading volume}}{\mbox{open interest}}.
\end{align} 
  Intuitively,  total trading volume is a proxy for speculation activity   and hedging demand is represented by   open interest.  A lower
ratio  implies lower speculative
activity relative to hedging demand, or vise versa. 

Second, we estimate leverage using liquidation data. At time $t$, suppose the futures price is $F_t$ (USD) and an investor enters one long (short) position with leverage ${\mathcal{L}}_{long}$ (${\mathcal{L}}_{short}$).  {The margin required in XBT is $\frac{1}{\mathcal{L}_{long}} \frac{1}{F_{t}}$ for long position and $\frac{1}{\mathcal{L}_{short}} \frac{1}{F_{t}}$ for short position. } According to long and short pay-offs in \eqref{payoff-long} and \eqref{payoff-short}, at a later time $s$, margin call is triggered once the trading loss wipes out margin deposit, i.e., 
\begin{align}\label{margin}
	\frac{1}{F_s} - \frac{1}{F_{t}} = \frac{1}{{\cal L}_{long}} \frac{1}{F_t}, \quad \mbox{and} \quad
	\frac{1}{F_t} - \frac{1}{F_{s}} = \frac{1}{{\cal L}_{short}} \frac{1}{F_t}.
\end{align}

As Coinaylze does not report the exact time when each forced liquidation happens, we made further assumption that investors open their positions at the opening  of day $d$ and the possible forced liquidation happens when the perpetuals price hits the daily low (high) for the long (short) position on day $d$. The corresponding returns are:
	\begin{align}\label{return-max}
		r_d^{max} = \frac{F_d^{high}- F_d^{open}}{F_d^{open}}, \quad \mbox{and} \quad
		r_d^{min} = \frac{F_d^{low}- F_d^{open}}{F_d^{open}}.
	\end{align}
	Here, $F_d^{open}$, $F_d^{high}$ and $F_d^{low}$ stand for the open, high, and  low prices. 
	 Once got forced liquidation, we could    infer traders' minimal long and short leverages  via \eqref{margin} as
\begin{align}\label{leverage}
	{\cal L}_{long} = - \frac{1+r_d^{min}}{r_d^{min}}, \quad \mbox{and} \quad
	{\cal L}_{short} =  \frac{1+r_d^{max}}{r_d^{max}}.
\end{align}
The long (short) liquidation percentage   $p_{long}$ ($p_{short}$) is estimated by the proportion of long (short) liquidation to open interest.

\begin{center}
	$<$Insert Table \ref{liqu_lev-summary} here.$>$
\end{center}

\begin{figure}[htbp]
	\centering
	\begin{minipage}[b]{1.1\textwidth}
		\includegraphics[width=1\textwidth]{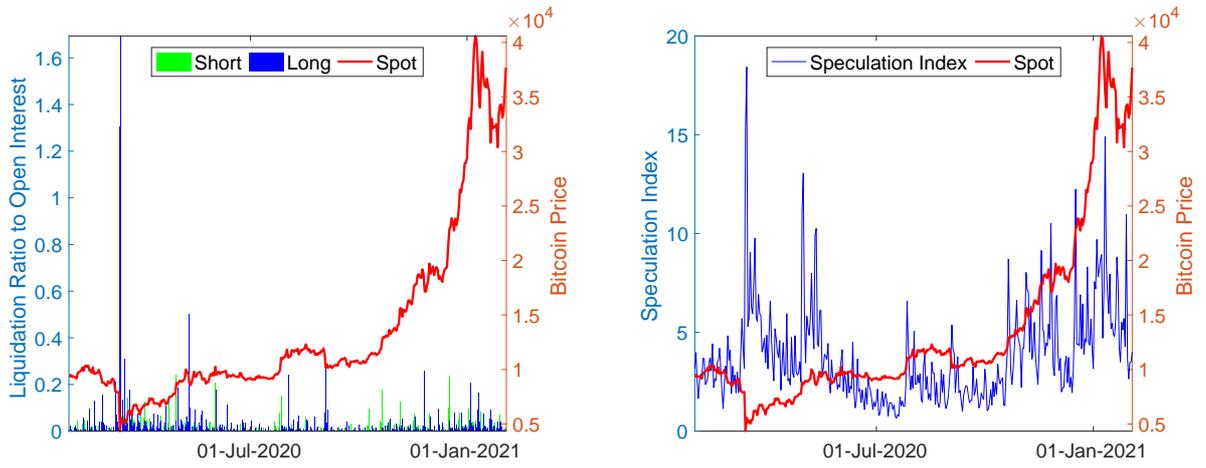}
	\end{minipage}
	\caption{Liquidation Ratio and Speculation Index}
	\label{speculation}
	\floatfoot{Note. The speculation index is defined as  the ratio  of     trading volume to   open interest, $
		\cal{SI} = \frac{\mbox{trading volume}}{\mbox{open interest}}.$   A lower
		ratio  implies lower speculative
		activity relative to hedging demand.  }
\end{figure}

\begin{center}
	$<$Insert Figure \ref{speculation} here.$>$
\end{center}

The summary statistics for maximal and minimal futures returns, speculation index, liquidation percentage,   and leverages are presented in Table \ref{liqu_lev-summary}. The average of speculation index sits at 3.75 which is much larger than S\&P 500 (0.15), Nikkei (0.21) and DAX (0.45) in \cite{lucia2010measuring}. Figure \ref{speculation} plots its time series in the sample period and shows more speculative trades in both bullish and bearish markets.   This reflects  extreme   speculative activity in bitcoin futures market. 
The daily average percentages of long and short liquidation to open interest are 3.51\% and 1.89\%, and the average leverage used by those \textit{got forced liquidation} is at least 58.13X for long and 59.94X for short. This shows liquidation is substantial due to high spot price risk and  some investors'  aggressive  leverage. 

\section{Conclusion}\label{4}
It is crucial and of self-interest for exchanges keeping futures margins at a proper level: high enough to preserve market integrity yet low enough to attract broad participants and maintain market liquidity. This paper employed the generalized extreme value theory to address optimal margins for bitcoin futures. Using the BitMEX perpetual futures, we found: first,  average leverages used by forced liquidation traders are  58.13X for long and 59.94X for short for the short position. Second, high leverage leads to substantial forced liquidation, where the daily liquidation percentage is 3.51\%   for long and 1.89\% for short. Third, investors should be more rational and cautious to use leverage in the concurrent high volatile market. We suggest  5X leverage for short and 3X leverage for long if one accepts 1\% daily margin call probability. Furthermore, the normality assumption of return will significantly underestimate margin levels by at least half.

The policy implications of our results are: (1) exchanges should give more clear documentation of the asymmetric risk character of perpetuals to market participants;  (2) exchanges should also consider perpetuals' inverse-payoff structure and set higher margin for long positions than shorts; (3) it is necessary and important for exchanges to limit leverage allowed and impose more margin deposit to lower forced liquidation, maintain market integrity,   and improve the functioning of futures markets. 

\cleardoublepage \newpage
\bibliographystyle{apalike}
\bibliography{option}

\cleardoublepage \newpage
\noindent\textbf{\Large Appendix}

\section*{Figures}
\newpage 
\newpage

\newpage

\newpage

\section*{Tables}
\begin{table}[htbp]\centering
	\resizebox{\textwidth}{!}{
		\caption{Summary Statistics of  Price Changes $\Delta F$ and $\Delta \widehat{F}$ on BitMEX}
		\label{bitmex_summary}
		\begin{tabular}{cccccccccccc}
			\toprule
			Variable   & \multicolumn{5}{c}{Standard Futures $\Delta F$}                          & \multicolumn{5}{c}{Perpetuals $\Delta \widehat{F}$}          \\
			\midrule
			$ t$          & 5min    & 30min   & 1h      & 8h      & 1d  &    & 5min    & 30min   & 1h      & 8h      & 1d      \\
			\midrule
			Min    & $-$0.2230  & $-$0.1671   & $-$0.2514    &$ -$0.2902    & $-$0.4526   & & $-$0.2870    & $-$0.2006    & $-$0.3358     & $-$0.4088     & $-$0.8267     \\
			P25    & $-$0.0008 & $-$0.0020  & $-$0.0027    & $-$0.0079    &$ -$0.0146    & & $-$0.0008   & $-$0.0020   &$ -$0.0027      & $-$0.0080    & $-$0.0148      \\
			Median  & 0.0000 & 0.0001  & 0.0001      & 0.0008     & 0.0027      & &0.0000     & 0.0001      & 0.0001       & 0.0008       & 0.0027       \\
			Mean       & 0.0000   & 0.0001  & 0.0002 & 0.0012 & 0.0035  &    & 0.0000     & 0.0000   & 0.0001     & 0.0005        & 0.0014                 \\
			P75     & 0.0009   & 0.0022      & 0.0031   & 0.0108  & 0.0223         &       & 0.0009   & 0.0022       & 0.0031     & 0.0106        & 0.0218   \\
			Max  & 0.3190  & 0.2396    & 0.1440   & 0.2333    & 0.2820      &       & 0.2419       & 0.1933         & 0.1259  & 0.1892           & 0.2200      \\
			Skewness    & 1.5741   & 0.0689    & $-$0.5213   & $-$0.0481    & $-$0.2415 &  & $-$1.6286 & $-$1.1195   & $-$1.9050  & $-$1.3412    & $-$3.6652     \\
			Kurtosis & 434.28   & 61.44    & 39.27   & 15.35   & 14.10       &          & 358.63                & 59.91                 & 68.47                 & 24.75                 & 65.36                  \\
			S.D.                  & 0.0030                & 0.0069                & 0.0096                & 0.0253                & 0.0446    &            & 0.0030                & 0.0069                & 0.0097                & 0.0257                & 0.0476                 \\
			Nobs                  & 431346                & 71891                 & 35945                 & 4493                  & 1497      &            & 431346                & 71891                 & 35945                 & 4493                  & 1497                   \\
			Stationary            & Yes                   & Yes                   & Yes                   & Yes                   & Yes    &               & Yes                   & Yes                   & Yes                   & Yes                   & Yes     \\     
			\bottomrule
	\end{tabular}}
	\floatfoot{Note. It reports summary statistics of price changes  $\Delta F$ and $\Delta \widehat{F}$ using BitMEX perpetual futures from January 1, 2017, to February 6, 2021, sampled at 5min, 30min, 1h, 8h, and 1d frequencies. P25 and P75 refer to 25\% and 75\% quantiles, and S.D. is the standard deviation. }
\end{table}

\begin{table}[htbp]
	\resizebox{\textwidth}{!}{
		\caption{GEV Parameter  Estimations   for $\Delta F$ and $\Delta \widehat{F}$}
		\label{para-bitmex}
		\begin{tabular}{cccccccccccc}
			\toprule
			Variable           & \multicolumn{5}{c}{Standard Futures $\Delta F$}                                & \multicolumn{5}{c}{Perpetuals $\Delta \widehat{F}$}                      \\
			\midrule
			$t$                  & 5min     & 30min    & 1h       & 8h       & 1d   &    & 5min     & 30min    & 1h       & 8h       & 1d       \\
			\midrule
			\multicolumn{11}{l}{Panel A: Right Tail }                                                                                     \\
			\multirow{2}{*}{$\tau$} & 0.3939   & 0.3643   & 0.3163   & 0.2386   & 0.2097   && 0.3845   & 0.3455   & 0.2904   & 0.1948   & 0.1424   \\
			& (0.0147) & (0.0268) & (0.0362) & (0.0517) & (0.0731) && (0.0147) & (0.0267) & (0.0359) & (0.0507) & (0.0713) \\
			\multirow{2}{*}{$\sigma$} & 0.2967   & 0.6119   & 0.9078   & 1.7176   & 2.8115   & &0.2938   & 0.6003   & 0.8817   & 1.6229   & 2.5662   \\
			& (0.0046) & (0.0163) & (0.0330) & (0.0929) & (0.2127) && (0.0045) & (0.0158) & (0.0316) & (0.0857) & (0.1877) \\
			\multirow{2}{*}{$\mu$} & 0.4281   & 0.8947   & 1.3809   & 2.7411   & 4.5184   & &0.4265   & 0.8879   & 1.3642   & 2.6731   & 4.3345   \\
			& (0.0051) & (0.0185) & (0.0387) & (0.1141) & (0.2648) && (0.0051) & (0.0182) & (0.0376) & (0.1078) & (0.2415) \\
			\midrule
			\multicolumn{11}{l}{Panel B: Left Tail }                                                                                      \\
			\multirow{2}{*}{$\tau$} & 0.4389   & 0.4112   & 0.3326   & 0.2641   & 0.2261  & & 0.4490   & 0.4311   & 0.3620   & 0.3152   & 0.3102   \\
			& (0.0156) & (0.0278) & (0.0391) & (0.0568) & (0.0785) && (0.0156) & (0.0278) & (0.0392) & (0.0577) & (0.0811) \\
			\multirow{2}{*}{$\sigma$} & 0.3057   & 0.6173   & 0.9149   & 1.7020   & 2.6053  & & 0.3087   & 0.6294   & 0.9414   & 1.7928   & 2.8165   \\
			& (0.0049) & (0.0170) & (0.0342) & (0.0954) & (0.2027) && (0.0050) & (0.0175) & (0.0358) & (0.1033) & (0.2293) \\
			\multirow{2}{*}{$\mu$} & 0.4276   & 0.8704   & 1.3417   & 2.4324   & 3.6979   && 0.4291   & 0.8768   & 1.3568   & 2.4840   & 3.8108   \\
			& (0.0053) & (0.0188) & (0.0396) & (0.1149) & (0.2491) && (0.0054) & (0.0191) & (0.0407) & (0.1210) & (0.2695) \\
			\midrule
			\multicolumn{11}{l}{Panel C: Common Tail  }                                                                                    \\
			\multirow{2}{*}{$\tau$} & 0.4165   & 0.3876   & 0.3237   & 0.2447   & 0.2096   && 0.4176   & 0.3901   & 0.3272   & 0.2531   & 0.2255   \\
			& (0.0107) & (0.0193) & (0.0266) & (0.0380) & (0.0528) && (0.0107) & (0.0193) & (0.0265) & (0.0378) & (0.0511) \\
			\multirow{2}{*}{$\sigma$} & 0.3012   & 0.6149   & 0.9121   & 1.7242   & 2.7480   && 0.3012   & 0.6148   & 0.9118   & 1.7197   & 2.7270   \\
			& (0.0033) & (0.0118) & (0.0238) & (0.0668) & (0.1480) && (0.0033) & (0.0118) & (0.0238) & (0.0668) & (0.1470) \\
			\multirow{2}{*}{$\mu$} & 0.4278   & 0.8825   & 1.3616   & 2.5891   & 4.1036   && 0.4277   & 0.8819   & 1.3601   & 2.5805   & 4.0740   \\
			& (0.0037) & (0.0132) & (0.0277) & (0.0816) & (0.1840) && (0.0037) & (0.0132) & (0.0277) & (0.0812) & (0.1815)\\
			
			\bottomrule
	\end{tabular}}
	\floatfoot{Note: This table reports left and right tail parameter estimations for  $\Delta F$ and $\Delta \widehat{F}$ under 5min, 30min, 1h, 8h, and 1d observation frequencies. For comparison purposes, we take the absolute value of the left tail to estimate corresponding extreme parameters and report a common tail that simultaneously encompasses absolute left tail and right tail.  Since variables $\Delta F_{t}$ and $\Delta \widehat{F}_{t}$ are quite small, we multiply them by 100. The corresponding standard errors are included in parentheses. }
\end{table}

\newpage

\begin{table}[h!]
	\caption{Optimal Margins for Bitcoin Standard Futures and Perpetual Futures}
	\label{oml-bitmex}
	\begin{tabular}{cccccccccc}
		\toprule
		& \multicolumn{4}{c}{Standard Futures Margin (\%)}         &  & \multicolumn{4}{c}{Perpetual Futures Margin (\%)} \\
		\midrule
		Margin Call Probability            & 0.1    & 0.05   & 0.01    & 0.001   &  & 0.1    & 0.05   & 0.01    & 0.001   \\
		\midrule
		\multicolumn{10}{l}{Panel A: Short Position}                                                          \\
		\multirow{2}{*}{5min}  & 1.50   & 2.10   & 4.29    & 11.12   && 1.48   & 2.06   & 4.14    & 10.54   \\
		& (0.38) & (0.49) & (0.69)  & (0.92)  && (0.38) & (0.49) & (0.69)  & (0.92)  \\
		\multirow{2}{*}{30min} & 3.03   & 4.17   & 8.19    & 20.01   && 2.93   & 4.00   & 7.67    & 18.05   \\
		& (0.88) & (1.13) & (1.60)  & (2.13)  && (0.89) & (1.14) & (1.61)  & (2.14)  \\
		\multirow{2}{*}{1h}    & 4.36   & 5.85   & 10.81   & 24.02   && 4.16   & 5.52   & 9.88    & 20.89   \\
		& (1.21) & (1.56) & (2.21)  & (2.94)  && (1.23) & (1.59) & (2.24)  & (2.98)  \\
		\multirow{2}{*}{8h}    & 7.86   & 10.17  & 17.12   & 32.95   && 7.26   & 9.20   & 14.76   & 26.34   \\
		& (3.13) & (4.05) & (5.77)  & (7.70)  && (3.24) & (4.17) & (5.92)  & (7.88)  \\
		\multirow{2}{*}{1d}    & 12.60  & 16.11  & 26.29   & 48.19   && 11.14  & 13.82  & 21.01   & 34.50   \\
		& (5.37) & (6.99) & (10.02) & (13.43) && (5.96) & (7.69) & (10.94) & (14.58) \\
		\midrule
		\multicolumn{9}{l}{Panel B: Long Position}                                                  &        \\
		\multirow{2}{*}{5min}  & 1.60   & 2.30   & 4.98    & 14.17   && 1.63   & 2.35   & 5.17    & 15.03   \\
		& (0.38) & (0.49) & (0.69)  & (0.92)  && (0.38) & (0.49) & (0.69)  & (0.92)  \\
		\multirow{2}{*}{30min} & 3.16   & 4.46   & 9.32    & 25.07   && 3.27   & 4.67   & 10.03   & 28.10   \\
		& (0.89) & (1.14) & (1.61)  & (2.14) & & (0.89) & (1.14) & (1.61)  & (2.14)  \\
		\multirow{2}{*}{1h}    & 4.41   & 5.98   & 11.30   & 25.96   && 4.63   & 6.38   & 12.50   & 30.45   \\
		& (1.24) & (1.59) & (2.24)  & (2.97)  && (1.25) & (1.60) & (2.26)  & (3.00)  \\
		\multirow{2}{*}{8h}    & 7.66   & 10.11  & 17.71   & 35.93   && 8.36   & 11.30  & 21.05   & 46.98   \\
		& (3.36) & (4.28) & (6.00)  & (7.93)  && (3.34) & (4.27) & (6.02)  & (7.99)  \\
		\multirow{2}{*}{1d}    & 11.34  & 14.73  & 24.78   & 47.11   && 12.98  & 17.55  & 32.56   & 72.10   \\
		& (6.06) & (7.68) & (10.72) & (14.12) && (6.24) & (7.97) & (11.22) & (14.86) \\
		\midrule
		\multicolumn{9}{l}{Panel C: Common Position}                                                &         \\
		5min                   & 1.55   & 2.20   & 4.62    & 12.55  & & 1.55   & 2.20   & 4.63    & 12.61   \\
		30min                  & 3.09   & 4.31   & 8.73    & 22.37  & & 3.10   & 4.33   & 8.79    & 22.63   \\
		1h                     & 4.38   & 5.91   & 11.03   & 24.90   && 4.39   & 5.94   & 11.13   & 25.28   \\
		8h                     & 7.76   & 10.12  & 17.26   & 33.74  & & 7.80   & 10.19  & 17.55   & 34.81   \\
		1d                     & 12.55  & 15.88  & 25.03   & 42.91   && 12.07  & 15.61  & 26.10   & 49.39  \\
		\bottomrule
	\end{tabular}
	\floatfoot{Note: This table reports optimal margins for Bitcoin standard futures and perpetual futures for short (Panel A) and long (Panel B) positions. The optimal margins for long and short positions are calculated via \eqref{long-margin-gev} and \eqref{short-margin-gev}. Panel C gives the margin level for the common position. The corresponding optimal margin levels obtained under the assumption of normality are also presented in parentheses, which are calculated via $p^{long}=\frac{1}{\sqrt{2 \pi} \sigma} \int_{-\infty}^{-ML^{long}} e^{-(x-\mu)^{2} / 2 \sigma^{2}}dx$ \mbox{and}	$p^{short}=\frac{1}{\sqrt{2 \pi} \sigma} \int_{M L^{\text {short }}}^{+\infty} e^{-(x-\mu)^{2} / 2 \sigma^{2}}dx$. Here, $\mu$ and $\sigma$ are the mean and standard deviation of   price changes $\Delta F_{t}$ and $\Delta \widehat{F}_{t}$. $ML_{long}$ and $ML_{short}$ denote optimal margins of long and short positions.
	}
\end{table}

\newpage

\begin{table}[htbp]\centering
	\caption{Summary Statistics for Speculation, Liquidation and Leverage}
	\setlength{\tabcolsep}{2mm}{
		\label{liqu_lev-summary}
		\begin{tabular}{cccccccccc}
			\toprule
			Variables   &$r_d^{min}$& $r_d^{max}$      & long liq.(M) & short liq.(M)&  $\cal{SI}$  & $p_{long}(\%)$ & $p_{short}(\%)$ & ${\cal L}_{long}$ & ${\cal L}_{short}$  \\
			\midrule
			min    & -0.46 & 0.00 & 0.00   & 0.00   & 0.66  & 0.00   & 0.00  & 1.16   & 3.84   \\
			median & -0.02 & 0.02 & 6.93   & 4.60   & 3.12  & 1.22   & 0.86  & 55.97  & 56.34  \\
			mean   & -0.03 & 0.03 & 20.14  & 10.17  & 3.75  & 3.51   & 1.89  & 58.13  & 59.94  \\
			max    & 0.00  & 0.35 & 843.39 & 132.70 & 18.42 & 169.39 & 24.32 & 100.00 & 100.00 \\
			Nobs   & 372   & 372  & 372    & 372    & 372   & 372    & 372   & 372    & 372          \\
			\bottomrule
	\end{tabular}}
	\floatfoot{Note. The maximal and minimal return $r_d^{min}$ and $r_d^{max}$ are estimated via \eqref{return-max}. The speculation index is defined as  the ratio  of     trading volume to   open interest, $
		\cal{SI} = \frac{\mbox{trading volume}}{\mbox{open interest}}.$ The long and short liquidation percentage $p_{long}$ and $p_{short}$ are estimated by the proportion of long/short liquidation to open interest. The long and short leverage ${\cal L}_{long}$ and ${\cal L}_{short}$ are estimated via \eqref{leverage}. All the data is daily frequency from January 29, 2020 to February 3, 2021. M denotes the Million USD.}
\end{table}

\end{document}